\documentstyle[preprint,pre,aps,epsbox]{revtex}

\newcommand{\noi}{\noindent}

\newcommand{\beq}{\begin{equation}}
\newcommand{\beqa}{\begin{eqnarray}}
\newcommand{\eeq}{\end{equation}}
\newcommand{\eeqa}{\end{eqnarray}}

\newcommand{\sekibun}[4]{\displaystyle \int_{#1}^{#2} #3 d #4}

\newcommand{\bubun}[2]{\frac{\partial #1}{\partial #2}}
\newcommand{\bibun}[2]{\frac{d #1}{d #2}}

\begin{document}

\title{Irreversibility resulting from contact with a heat bath caused
by the finiteness of the system}

\draft

\author{Katsuhiko Sato and Ken Sekimoto}
\address{Yukawa Institute for Theoretical Physics, Kyoto University,
Kyoto 606-8502, Japan
}
\author{Tsuyoshi Hondou}
\address{Department of Physics, Tohoku University,
Sendai 980-8578, Japan
}
\author{Fumiko Takagi}
\address{Satellite Venture Business Laboratory,
        Ibaraki University,
        Ibaraki 316-8511, Japan
}
\date{\it 27/07/00}
\maketitle

\begin{abstract}
When a small dynamical system that is initially in contact with a heat
bath
is detached from this heat bath and then caused to undergo a
quasi-static adiabatic processes, the resulting statistical
distribution of the system's energy differs from that
of an equilibrium ensemble. Subsequent contact of the system
with another heat bath is inevitably irreversible, hence
the entire process cannot be reversed without a net energy transfer
to the heat baths.
\end{abstract}


\section{Introduction}
\label{sec:intro}
Ordinary thermodynamics assumes the extensivity of the system
in question, and it is not applicable directly to finite systems.
Hill\cite{hill} developed a framework to deal with
systems which are moderately large and homogeneous, except
for their boundaries.
In this framework, the corrections to the thermodynamic behaviour
due to the the effect of the surfaces and the edges of the system
are incorporated in the form of an expansion in the number of
the constituent atoms, $N$.
Our interest here is in systems further removed from the
thermodynamic limit, such as
mesoscopic devices and molecular motors,
which are intrinsically small
and heterogeneous and for which the method of Ref.\cite{hill} is
not sufficient.
Hereafter, we call such systems `small systems'.

\begin{figure}[ht]
\vspace{2mm}
\hspace{5mm}
\postscriptbox{15cm}{6.0cm}{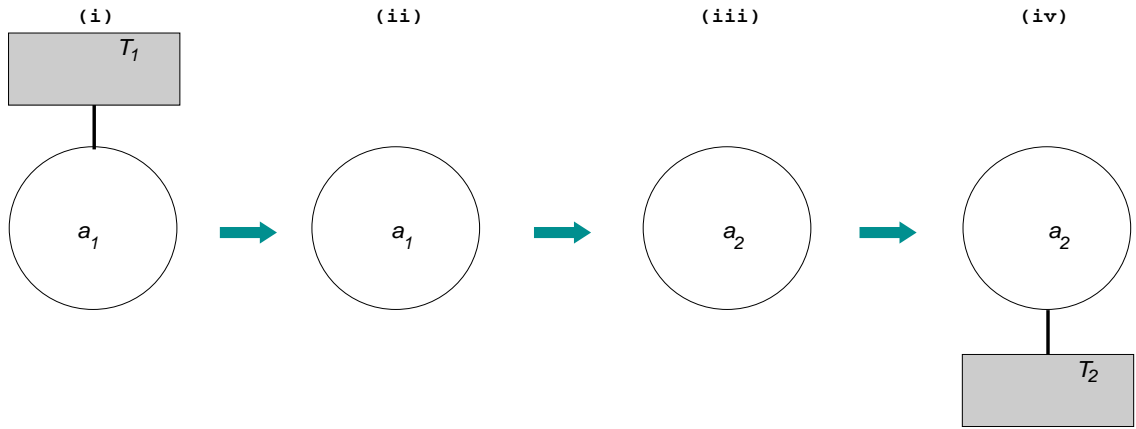}
\vspace{0cm}
\caption{The process $\{T_1,a_1;T_2,a_2\}$ is
schematically depicted
as (i)$\Rightarrow$(ii)$\Rightarrow$(iii)$\Rightarrow$(iv).
The gray boxes represent the heat baths
at the temperatures indicated therein, and the circles represent
the small system. The thick solid lines in (i) and (iv) denote
the thermal contact between the small system and
the two heat baths.}
\label{fig:Fig1}
\end{figure}

In this paper our purpose is to elucidate the distinctive
nature of small systems by considering
the following process, which we denote by
$\{T_1,a_1;T_2,a_2\}$  (See Fig.~\ref{fig:Fig1}).
%
\begin{description}
\item[(i)] First, a small system is in thermal contact with a heat
bath of temperature $T_1$. (Throughout this paper we assume that both
the interaction energy associated with the thermal contact and the
work required to change this contact are negligibly small
\cite{molecular-carnot}.)  \item[(ii)] We then gradually remove the
thermal contact between the system and the heat bath.  \item[(iii)]
Next, we change some arbitrary control parameter of the system, $a$,
from its initial value $a_1$ to a new value $a_2$ {\it
quasi-statically}.  We measure the work required to make this change
as the increase of the energy of the small system.
\item[(iv)] Finally, we gradually establish a thermal contact between
the system and

the second heat bath of temperature $T_2$.
\end{description}
We now introduce the concept of the `reversibility' associated with
the process $\{T_1,a_1;T_2,a_2\}$.
\indent\indent \begin{minipage}[t]{15cm} {\underline{\it Definition}:
 The process $\{T_1,a_1;T_2,a_2\}$ is called ``reversible'' if no net
 energy is transfered, on the statistical average over infinite number
 of repetitions, from or to either heat bath through the composite
 processes of $\{T_1,a_1;T_2,a_2\}$ followed by
 $\{T_2,a_2;T_1,a_1\}$.}  If the process $\{T_1,a_1;T_2,a_2\}$ is not
 reversible, it is called ``irreversible.''
\end{minipage}\\

\noi Reversibility, therefore, implies that the statistical average of
the work needed for the process $\{T_1,a_1;T_2,a_2\}$ is the opposite
of that for the process $\{T_2,a_2;T_1,a_1\}$.
In macroscopic systems, reversibility holds if and only if $T_2$ is
equal to the temperature of the (macroscopic) system after operation
(iii).  This fact is a prerequisite for the existence of
thermodynamics, in which the Helmholtz free energy can be used to
relate equilibrium states at different temperatures.  For small
systems, however, the situation is completely different:\\
\indent\indent \begin{minipage}[t]{15cm} {\underline{\it Statement}:
The processes $\{T_1,a_1;T_2,a_2\}$ for small systems are
irreversible, except for some ``special'' cases.}
\end{minipage}
\null \\

It is important to note that for small systems we cannot define the
temperature unambiguously, at least when they are isolated, and the
energy of the system at the end of operation (ii) is a strictly
statistical quantity. (This is related to the fact that the operation
of removing the thermal contact is intrinsically irreversible, however
small the work associated with this operation.)
In order to understand intuitively how these features of small systems
lead to irreversibility, we first describe qualitatively what happens
in the processes (i)-(iv).

In (i), the energy $E$ of the small system fluctuates, and its
statistics obey the canonical ensemble at temperature $T_1$.  In (ii),
the energy of the system is fixed at a particular value.  This energy
$E$ is a stochastic variable, and its distribution is given by the
canonical ensemble at temperature $T_1$, as long as the removal of the
thermal contact with the heat bath is sufficiently
gentle\cite{molecular-carnot}.  In (iii), the energy of the small
system changes in such a manner that the phase volume enclosed by a
constant energy surface, $J(E,a)$, ( { see (\ref{eq:J-definition}) in
the text } ) is invariant. This follows from the ergodic invariant
theorem \cite{kasuga}.  With the exception of those systems for which
$J(E,a)$ has a special functional property, the statistical
distribution of $E$ at the end of this adiabatic process is no longer
consistent with the canonical ensemble at {\it any} temperature.  In
(iv), this non-canonical distribution of the energy relaxes {\it
irreversibly} (in the ordinary sense) to the canonical distribution at
the temperature $T_2$, whether or not, on statistical average, the net
energy transfer between the system and the heat bath is zero.  We note
that the essential feature distinguishing small systems from
macroscopic systems is the distortion of the energy distribution in
(iii), which can be neglected in macroscopic thermodynamics.

In the following sections we prove the above statement with an
argument based on the ergodicity hypothesis of Hamiltonian dynamical
systems.  The outline of the proof is as follows.  In
\S~\ref{seclemmas} we prove the three lemmas as preparatory steps for
the main statement. In \S~\ref{secstatement} we prove the main
statement. In \S~\ref{secdiscussion} we discuss the physical meaning
of the Statement. We also show there the necessary condition for the
process $\{T_1,a_1;T_2,a_2\}$ to be reversible.

\section{Three Lemmas}
\label{seclemmas}
In order to prove the Statement, we first introduce three lemmas.

{\em Lemma 1} ---
The entropy $S$ (see below) remains invariant in the process
(iii).

{\em Proof:}  We consider an ensemble of the mechanical system which is
described by a time dependent Hamiltonian $H$.  Let us denote by
$P(\Gamma,t)$ the normalized distribution function of the ensemble at
time $t$, where $\Gamma$ is the phase coordinates of the system, i.e.,
the position coordinates and the momenta of the system.  The entropy
$S$ is defined as, a functional of the normalized distribution $P$,
\begin{equation}
S[P(\cdot,t)] \equiv -\int_{\{\Gamma\}} P(\Gamma,t) \log P(\Gamma,t) d
\Gamma,
\label{eq:01}
\end{equation}
where the symbol $\{\Gamma\}$ indicates that the integral is taken
over the whole phase space.

Let us now examine the behaviour of the entropy $S$ with time. First,
we note that the time evolution of the distribution function is
described by the so-called Liouville's equation:
\begin{eqnarray}
\bubun{P(\Gamma,t)}{t} &=& - \left( \bubun{}{q} \bubun{H(\Gamma,t)}{p}
- \bubun{}{p} \bubun{H(\Gamma,t)}{q} \right) P(\Gamma,t) \nonumber \\
&\equiv& - \bubun{}{\Gamma} [ V_{(\Gamma,t)} P(\Gamma,t) ] \nonumber
\label{eq:02},
\end{eqnarray}
where $q$ and $p$ are the position coordinates and the momenta of the
system, respectively, and we have introduced here the velocity of the
system point in the phase space, $V_{(\Gamma,t)}$.  The velocity
$V_{(\Gamma,t)}$ satisfies
\begin{equation}
\bubun{}{\Gamma} V_{(\Gamma,t)}=0
\label{eq:03}
\end{equation}
which can be checked by the equation of motion of the mechanical
system.

Using (
\ref{eq:03}
), we evaluate the time derivative of the entropy
$S$:
\begin{eqnarray*}
\bubun{S[P(\cdot,t)]}{t}
&=&
- \int_{\{\Gamma\}}
\bubun{P(\Gamma,t)}{t} \left.  \bibun{(x \log x)}{x}
\right|_{x=P(\Gamma,t)} d\Gamma \nonumber \\
&=&
 \int_{S_{\Gamma}} V_{(\Gamma,t)} P(\Gamma,t) \log P(\Gamma,t)
dS_{\Gamma },
\end{eqnarray*}
where the symbol $S_{\Gamma}$ represents the surface integral over the
surface enclosing the phase space and we have used (\ref{eq:03}) and
performed the integration by parts.  As we are interested in a
mechanical system such that all particles are confined in a finite
region in position space and that the Hamiltonian involves the kinetic
energy terms $p^2/2 m$, $P(\Gamma,t)$ vanishes at any point on
$S_{\{\Gamma\}}$.
The lemma applies to the process (iii), since a quasi-static adiabatic
process is realized by a time-dependent Hamiltonian.

{\em Lemma 2} --- A canonical distribution is the distribution to
maximize the entropy $S$ subject to the constraint that the ensemble
average of the energy is $E$, i.e.,
\begin{equation}
\int_{\{\Gamma\}} H(\Gamma) P(\Gamma) d\Gamma =E,
\label{eq:04}
\end{equation}
where the canonical distribution characterized by the Hamiltonian $H$
and the temperature $T$ is defined as
\begin{equation}
P_c(\Gamma;T,H) \equiv \frac{e^{-\frac{H(\Gamma)}{T}}}{ Z(T,H) },
\label{eq:05}
\end{equation}
with $Z(T,H)$ being
the normalization constant,
\begin{equation}
Z(T,H) = \int_{\{\Gamma\}} e^{-\frac{H(\Gamma)}{T}} d\Gamma .
\label{eq:Z}
\end{equation}
{\em Proof:} 
Let us examine the difference between the entropies of
two distributions, the canonical distribution $P_c$ and any other
distribution $P$, both being normalized and satisfying the constraint
condition (\ref{eq:04}).  We find
\begin{eqnarray}
S[P_c] - S[P] &=& \int_{\{\Gamma\}} P(\Gamma) \log
\frac{P(\Gamma)}{P_c(\Gamma)} d\Gamma,
\label{eq:06}
\end{eqnarray}
where we have used (\ref{eq:05}) and the conditions,
\begin{eqnarray*}
& & \int_{\{\Gamma\}} P_c(\Gamma) d\Gamma = \int_{\{\Gamma\}}
P(\Gamma) d\Gamma=1, \\ & & \int_{\{\Gamma\}} H(\Gamma) P_c(\Gamma)
d\Gamma = \int_{\{\Gamma\}} H(\Gamma) P(\Gamma) d\Gamma =E.
\end{eqnarray*}
We have written here the canonical distribution as $P_c(\Gamma)$ for
the simplicity of notation, though precisely it implies
$P_c(\Gamma;T,H)$ in our notation. The right-hand side of
(\ref{eq:06}) is known as the {\it relative entropy} and has been
known to be non-negative, as we easily demonstrate as follows:
\begin{eqnarray}
\int_{\{\Gamma\}} P(\Gamma) \log \frac{P(\Gamma)}{P_c(\Gamma)} d\Gamma
&=& \int_{\{\Gamma\}} P(\Gamma) \log \frac{P(\Gamma)}{P_c(\Gamma)}
d\Gamma - \int_{\{\Gamma\}} \left[ P(\Gamma) - P_c(\Gamma) \right]
d\Gamma \nonumber \\ &=& \int_{\{\Gamma\}} P_c(\Gamma) \left(
\frac{P(\Gamma)}{P_c(\Gamma)} \log \frac{P(\Gamma)}{P_c(\Gamma)} -
\frac{P(\Gamma)}{P_c(\Gamma)} +1 \right) d\Gamma \nonumber \\ &\geq&
0.
\label{eq:08}
\end{eqnarray}
The inequality in the last line follows from the fact that $x \log x -
x + 1 \geq 0$ for $x \geq 0$.  The equality holds if and only if
$x=1$, so that only the canonical distribution realizes the maximum
value of $S$.  Thus the lemma is proved.

{\em Lemma 3} --- Let $\langle H \rangle_{(T,H)}$ be the ensemble
average of the Hamiltonian $H$ over the canonical distribution
$P_c(\Gamma;T,H)$.  (Hereafter we shall denote, in general, the
canonical average using the distribution $P_c(\Gamma;T,H)$ by $\langle
\cdot \rangle_{(T,H)}$, that is, for an arbitrary physical quantity
$A$ defined on the phase space: $ \langle A \rangle_{(T,H)} \equiv
\int_{\{\Gamma\}} A(\Gamma) P_c(\Gamma;T,H) d\Gamma .  $ ) Then
$\langle H \rangle_{(T,H)}$ is monotonically increasing with $T$.  The
entropy $S[P_c(\cdot;T,H)]$ is also monotonically increasing with $T$.

{\em Proof:} 
Differentiating $\langle H \rangle_{(T,H)} $ with respect
to $T$, we obtain
\begin{equation}
\bubun{ \langle H \rangle_{(T,H)} }{T} = \frac{\langle ( H- \langle H
\rangle_{(T,H)} )^2 \rangle_{(T,H)} }{T^2}
\label{eq:09}.
\end{equation}
Since the value of $H$ is indeed distributed under the canonical
distribution, the right-hand side of (\ref{eq:09}) is positive.

Likewise, differentiating $S[P_c(\cdot;T,H)]$ with respect to $T$, we
obtain
\begin{eqnarray}
\bubun{S[P_c(\cdot;T,H)]}{T} &=& \frac{1}{T} \bubun{ \langle H
\rangle_{(T,H)} } {T}.
\label{eq:10}
\end{eqnarray}
We note that the temperature $T$ is positive in most physical
situations.  Indeed, for the Hamiltonian involving the kinetic terms,
$T$ must be positive to satisfy the normalization condition of the
canonical distribution.  Thus, the right-hand side of (\ref{eq:10}) is
positive.

\newpage

\section{Proof of the Statement}\label{secstatement}
Let us consider an ensemble of the small systems whose Hamiltonian is
$H_a$, where $a$ is a parameter controlled from the outside.  We shall
analyse the two processes for the ensemble, $\{T_1,a_1;T_2,a_2\}$ and
its inverse $\{T_2,a_2;T_1,a_1\}$ with given the values of the
temperature $T_1$ and parameters $a_1$ and $a_2$.  A temperature $T_2$
is to be determined so that the heat bath of the temperature $T_2$
receives no energy from the ensemble of the small systems during the
process (iv) of $\{T_1,a_1;T_2,a_2\}$.  First, we consider the process
$\{T_1,a_1;T_2,a_2\}$.  When detached form the heat bath of the
temperature $T_1$ (the process (ii)), the ensemble is the canonical
ensemble characterized by $T_1$ and $H_{a_1}$.  The ensemble average
of the energy, $\bar{E}_1$, is then given by
\begin{equation}
\bar{E}_1 = \langle H_{a_1} \rangle_{(T_1,H_{a_1})}.
\label{eq:11}
\end{equation}
When the parameter $a$ of the system is quasi-statically changed along
the process (iii), the distribution of the systems, in general,
changes.  The final distribution is uniquely determined by the
adiabatic theorem.  (We will not write down the explicit form of the
distribution, since our proof does not depend on the concrete form of
the distribution.)  We will write the distribution of the ensemble at
$a$ as $P_{a}(\Gamma;T_1,H_{a_1})$, where $T_1$ and $H_{a_1}$ are the
arguments reminding us of the fact that the ensemble at $a=a_1$ was
the canonical ensemble with $T_1$ and $H_{a_1}$.  By our definition,
$P_a$ at any temperature $T$ and for any value of $a$ satisfies
\begin{equation}
P_{a}(\Gamma;T,H_{a})=P_c(\Gamma;T,H_{a}).
\label{eq:12}
\end{equation}

According to Lemma 1, the entropy $S$ remains invariant during the
process (iii):
\begin{equation}
S[P_{a_1}(\cdot;T_1,H_{a_1})]=S[P_{a_2}(\cdot;T_1,H_{a_1})].
\label{eq:13}
\end{equation}
At the end of (iii) the ensemble average of the energy, $\bar{E}_2$,
is expressed as 
\begin{equation}
\bar{E}_2 = \sekibun{\{\Gamma\}}{}{H_{a_2}(\Gamma)
P_{a_2}(\Gamma;T_1,H_{a_1}) } {\Gamma}.
\label{eq:14}
\end{equation}
For the process (iv), we choose the temperature $T_2$ so that the
average energy of the ensemble does not change upon the contact with
the heat bath of the temperature $T_2$.  It is because our aim is to
know whether or not the process $\{T_1,a_1;T_2,a_2\}$ can be made
reversible. $T_2$ must, therefore, satisfy
\begin{equation}
\bar{E}_2 = \langle H_{a_2} \rangle_{(T_2,H_{a_2})}.
\label{eq:15}
\end{equation}
According to Lemma 2, the relations (\ref{eq:14}) and (\ref{eq:15})
imply
\begin{equation}
S[P_{a_2}(\cdot;T_1,H_{a_1})] \leq S[P_c(\cdot;T_2,H_{a_2})],
\label{eq:16}
\end{equation}
where the equality holds only if the ensemble at the end of (iii) is
the canonical ensemble.  If our system is such that the canonical
distribution is transformed into the canonical one through the process
(iii) of $\{T_1,a_1;T_2,a_2\}$, then it is also true for the process
(iii) of $\{T_2,a_2;T_1,a_1\}$, since (iii) is a quasi-static
adiabatic and is, therefore, reversible process.

Next, we examine the process $\{T_2,a_2;T_1,a_1\}$.  As above, the
process (iii) yields the relation
\begin{equation}
S[P_{a_2}(\cdot;T_2,H_{a_2})]=S[P_{a_1}(\cdot;T_2,H_{a_2})].
\label{eq:17}
\end{equation}
The ensemble average of the energy at the end of (iii), $\bar{E}_1'$,
is
\begin{equation}
\bar{E}_1' = \sekibun{\{\Gamma\}}{}{ H_{a_1}(\Gamma)
P_{a_1}(\Gamma;T_2,H_{a_2}) } {\Gamma}.
\label{eq:18}
\end{equation}
Now we ask if there is a non-zero flow of energy into the heat bath of
the temperature $T_1$ at the end of (iii) of the process
$\{T_2,a_2;T_1,a_1\}$ when we put the ensemble in contact with that
heat bath. To answer to this we only need to compare the value of
$\bar{E}_1'$ with that of $\bar{E}_1$ since the contact with the heat
bath forces the ensemble to obey the canonical distribution with the
average energy $\bar{E}_1$.  If $\bar{E}_1' > \bar{E}_1$, then the
positive energy, $\bar{E}_1' - \bar{E}_1$, flows from the ensemble to
the heat bath.

To see if this is the case, it is convenient to introduce the
temperature $T_1'$ which satisfies
\begin{equation}
\bar{E}_1' = \langle H_{a_1} \rangle_{(T_1',H_{a_1})},
\label{eq:19}
\end{equation}
that is, we temporally introduce the canonical ensemble whose the
ensemble energy is equal to $\bar{E}'$.  The equations (\ref{eq:18})
and (\ref{eq:19}) imply, with Lemma 2, that
\begin{equation}
S[P_{a_1}(\cdot;T_2,H_{a_2})] \leq S[P_c(\cdot;T_1',H_{a_1})].
\label{eq:20}
\end{equation}
Combining (\ref{eq:13}), (\ref{eq:16}), (\ref{eq:17}) and
(\ref{eq:20}), we arrive at the inequality
\begin{equation}
S[P_c(\cdot;T_1,H_{a_1})] \leq S[P_c(\cdot;T_1',H_{a_1})],
\label{eq:20.1}
\end{equation}
where we have used the property (\ref{eq:12}) of $P_a$.  According to
Lemma 3, this inequality (\ref{eq:20.1}) implies
$$
T_1 \leq T_1'
$$
and
\begin{equation}
\bar{E}_1 \leq \bar{E}_1' .
\label{eq:21}
\end{equation}
Thus we now complete the proof of the Statement: Given the temperature
$T_1$ and the parameters $a_1$ and $a_2$, no matter what we choose as
the temperature $T_2$, the process $\{T_1,a_1;T_2,a_2\}$ or
$\{T_2,a_2;T_1,a_1\}$ generally requires some non-negative energy to
move from the ensemble of the small systems to the heat baths. The
special case with no energy transfer is the reversible case as
mentioned below (\ref{eq:16}), that is, the only case that the
canonical distribution form of the ensemble is preserved along in the
quasi-static adiabatic process (iii).  We will discuss the condition
for this to occur in the next section.

\newpage

\section{Discussion }
\label{secdiscussion}

We first note that the inequality (\ref{eq:21}) is fundamental in the
sense that if it were violated, we could construct a perpetual machine
of the second kind with the following hypothetical protocol:

1.  We start from an ensemble of the small systems in contact with a
heat bath at temperature $T_1$.

2.  We detach these systems {\it gently} from the heat bath, and the
change the parameter $a$ from $a_1$ to $a_2$ quasi-statically. The
work necessary to make this change is $\bar{E}_2 - \bar{E}_1$ par
system.

3.  We now fix the parameter $a$ at $a_2$, and introduce the
interactions among these system. We assume that these interactions are
sufficiently smaller than the systemfs energy, but at the same time
large enough for the repartition of the energy within a certain time.

4.  We remove these interactions : the ensemble of the systems obeys
the canonical distribution characterized by $T_2$ and $H_{a_2}$. (Note
that we have not used any heat bath other than the initial one at the
temperature $T_1$.)

5.  We then slowly change the parameter $a$ from $a_2$ back to $a_1$.
The required work here is $ \bar{E}_1' - \bar{E}_2$.

6.  Finally, we close the cycle by bringing these small systems into
contact with the heat bath at temperature $T_1$.

If the inequality $ ( \bar{E}_2 - \bar{E}_1 ) +( \bar{E}_1' -
\bar{E}_2 ) <0$ were to hold in this cycle, we could obtain the
positive work $\bar{E}_1-\bar{E}_1'$ through the cycle, where the only
resource of the energy is the heat bath at temperature $T_1$.

\vspace{1cm} 

Below we will derive briefly the condition that the cycle of processes
discussed above become reversible.  This condition requires that the
distribution remains to be the canonical one upon quasi-static
adiabatic processes, see the paragraph below (\ref{eq:16}).  The
change of the distribution in those processes is governed by the
adiabatic theorem\cite{kasuga}: If we denote by $E_1$ and $E_2$ the
energy of the system before and after a quasi-static adiabatic
process, through which the parameter changes from $a_1$ to $a_2$,
respectively, the ``action'' $J(E,a)$ defined by
\begin{equation}
J(E,a)
\equiv
\int_{\{\Gamma\}}\theta(E-H_a(\Gamma))d\/\Gamma
\label{eq:J-definition}
\end{equation}
satisfies the following relationship:
\begin{equation}
J(E_1,a_1)=J(E_2,a_2).
\label{eq:JJ}
\end{equation}
Using (\ref{eq:JJ}) we can see how the energy distribution of the
system's ensemble changes through such process.  The energy
distribution before the process, $P(E_1)$, is given by construction as
\[ 
P(E_1)dE_1=\frac{e^{-\frac{E_1}{T_1}}}{Z(T_1,H_{a_1})}W(E_1,a_1)dE_1,
\]
where $Z$ has been defined below (\ref{eq:Z}) and $W(E,a)$ is defined
by $W(E,a)\equiv \frac{\partial J(E,a)}{\partial E}$.  Noting that
(\ref{eq:JJ}) and the above definition of $W(E,a)$ give
\[W(E_1,a_1)dE_1=W(E_2,a_2)dE_2,\]
the energy distribution after the process, $P'(E_2)$ is given as
\[
P'(E_2)dE_2=\frac{e^{-\frac{E_1}{T_1}}}{Z(T_1,H_{a_1})}W(E_2,a_2)dE_2.
\]

This distribution corresponds to the canonical one at some
temperature, say $T_2$, if and only if
\[\frac{E_1}{T_1}=\frac{E_2}{T_2}\]
is satisfied.  Thus, we reach the condition for the reversibility: the
adiabatic theorem (\ref{eq:JJ}) applied for a quasi-static adiabatic
process of the system should yield the relationship
\begin{equation}
E_2=\phi(a_1,a_2)E_1
\label{eq:rev-cond}
\end{equation}
with $\phi(a_1,a_2)$ being a function of the parameter values before
and after the process.

An example of the systems satisfying (\ref{eq:rev-cond}) is a harmonic
oscillator with the Hamiltonian, $ H_a= p^2/2 + a q^2/2.$ When the
spring constant $a$ is changed quasi-statically, the precess
$\{T_1,a_1;T_2,a_2\}$ is reversible.  By constrast, an example that
does not satisfy (\ref{eq:rev-cond}) is given by the following
Hamiltonian:
$$ H_a = p^2/2 + \exp \{ \frac{|q|}{a} \} $$
The proof, not shown here, is easy.

Our proof of (\ref{eq:21}) is for the systems obeying classical
dynamics.  After our work, H. Tasaki has shown that essentially the
same mechanism of irreversibility is found for the systems obeying
quantum mechanics \cite{tasaki}.  There, the proof has been done, just
we did here, using the fact that the canonical ensemble realizes the
maximum entropy among those ensembles with the same average energy.
We could say that it is this property of the canonical ensemble that
leads to the inequality (\ref{eq:21}).

In order to obtain a deeper physical insight of the inequality
(\ref{eq:21}), let us comparer the system which consists of infinitely
many small subsystems connected among each other with the system of
the ensemble of mutually isolated small systems.  We shall call the
these two systems the ``connected system'' and the ``disconnected
system'', respectively.  As the former system is macroscopic, we can
apply to it the ordinary thermodynamics and therefore the process
$\{T_1,a_1;T_2,a_2\}$ can be made reversible for such system.  To
assure it we must assume that the interaction energy assigned to the
coupling among the small subsystems is assumed to be ignorably small
while it is effective enough to attain the thermal equilibrium of the
whole connected system.  Under this assumption we can prove (not
shown) that the energy distribution of the small subsystems belonging
to the connected system remains to be the canonical one throughout the
process (iii).

\begin{figure}[hbtp]
\begin{center}
\epsfile{height=8cm,file=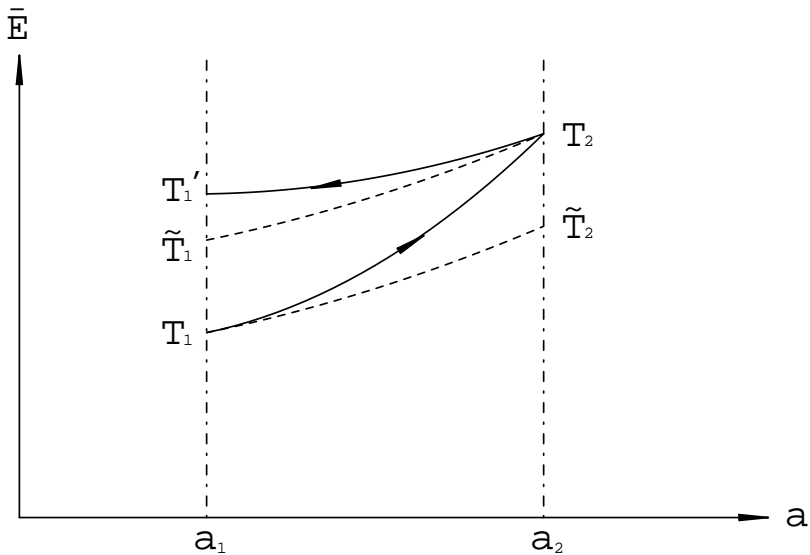} 
\caption{ {\bf Thick solid curves}: The average energy of the small
system, $\bar{E} $, as a function of the parameter $a$, along
quasi-static adiabatic processes. The arrows indicate the direction of
the processes. {\bf Dotted curves}: The energy of the combined system
per constituent small system along quasi-static adiabatic processes.
At each extreme point of the curves, the value of the average energy
is indicated by the corresponding temperature of the canonical
ensemble. For example, ${T}_1^{'}$ indicates that $ \bar{E}= \langle
H_{a_1} \rangle_{({T}_1^{'},H_{a_1})} $. At the point indicated by
$T_2$, the upper solid curve and the upper dotted curve are tangent,
and at the point indicated by $T_1$, the lower solid curve and the
lower dotted curve are tangent. }
\label{fig:Fig2}
\end{center}
\end{figure}

Furthermore the entropy related to this distribution, whose definition
has been given in ( \ref{eq:01}), is conserved during the process
(iii), as we can show easily by using the fact that the distribution
is kept to be canonical throughout this process.  That is, along the
process (iii) the canonical distribution $P_c(\Gamma;\tilde{T},H_a)$
of the connected system at the parameter value $a$ satisfies the
following relationship:
\[S[P_c(\cdot;{T_1},H_{a_1})]=S[P_c(\cdot;\tilde{T},H_a)].\]
This equality combined with Lemma 2 implies that, at any point along
the process (iii), the average energy of the small systems in the
disconnected system is generally not smaller than the average energy
of the small subsystems in the connected system (see
Fig.\ref{fig:Fig2} for the schematic illustration).  This figure
gives us the intuitive picture that the irreversibility of the
disconnected system is caused by its excess energy in reference to the
connective system which is reversible.

It is a future topic of investigation to determine if we can construct
a thermodynamic framework of small systems that can describe adiabatic
processes as well as isothermal processes for systems in contact with
heat baths. Our results imply that, in such framework, if there exists
a thermodynamic function whose difference calculated with respect to
two states is the quasi-static adiabatic work $\bar{E}_2-\bar{E}_1$ ,
then it cannot be the case that this function depends on only $T$ and
$a$.  (This is in contrast to the case of isothermal processes for a
small system in contact with a heat bath.  For such processes, using
the formalism of stochastic energetics\cite{ks1,ks-ss,ks2} it has been
shown that the Helmholtz free energy can be used to determine the work
necessary to move between two states by changing the value of $a$
sufficiently slowly that the small system evolves quasi-statically.)
To construct the thermodynamic framework of a small system, it is
desirable to find a method of characterizing in terms of work the
process through which the distribution changes from a non-canonical
form $P_{a_2}(\cdot;T_1,H_{a_1})$ to the canonical form
$P_c(\cdot;T_2,H_{a_2})$.  If this is possible, it is natural to
expect that the maximum of such extracted work to be $T_2
(S[P_c(\cdot;T_2,H_{a_2})]-S[P_{a_2}(\cdot;T_1,H_{a_1})])$ (see (
\ref{eq:16} )).
In any case, the quantity
$S[P_c(\cdot;T_2,H_{a_2})]-S[P_{a_2}(\cdot;T_1,H_{a_1})]$ is a strong
measure of the distance from the corresponding reversible process
since this is non-vanishing unless the functions
$P_{a_2}(\cdot;T_1,H_{a_1})$ and $P_c(\cdot;T_2,H_{a_2})$ are
identical.

\acknowledgements
Fruitful discussions with and comments from
T. Shibata,
S. Sasa,
G. Paquette,
A. Yoshimori,
and H. Tasaki are gratefully
acknowledged.
This work is supported in part by the COE scholarship (K. Sa.), a Grant
in Aid by the
Ministry of Education, Culture and Science (Priority Area,
No.11156216) (K.Se.) and by the Inamori Foundation (T.H.).


\clearpage


\begin{thebibliography}{99}
\bibitem{hill} T.L. Hill {\it Thermodynamics of Small Systems}
Parts I and II, (Dover New York 1994; Original publication from
Benjamin, 1963-1964).

\bibitem{molecular-carnot}
As for the details of the operation to realize this condition, see
K. Sekimoto, F. Takagi and T. Hondou, cond-mat/9904322.

\bibitem{kasuga} The general mathematical proof of the adiabatic
theorem can be found in:
T. Kasuga, Proc. Japan Academy, Tokyo, {\bf 37}, 366 (1961)
, {\it ibid.} 372, {\it ibid.} 377.

\bibitem{tasaki} H. Tasaki, cond-mat/0008420.

\bibitem{ks1}
K. Sekimoto, J. Phys. Soc. Jpn. {\bf 66}, 1234 (1997);

\bibitem{ks-ss}
K. Sekimoto and S. Sasa,  J. Phys. Soc. Jpn. {\bf 66}, 3326 (1997);

\bibitem{ks2}
K. Sekimoto: Progr. Theor. Phys. Suppl. {\bf 130}, 17 (1998).

\end{thebibliography}
\end{document}